\documentclass[sigplan,screen,authorversion]{acmart} 

\usepackage{xcolor}
\definecolor{ltblue}{rgb}{0,0.4,0.4}
\definecolor{dkblue}{rgb}{0,0.1,0.6}
\definecolor{dkgreen}{rgb}{0,0.35,0}
\definecolor{dkviolet}{rgb}{0.3,0,0.5}
\definecolor{dkpink}{rgb}{0.5,0,0.3}
\definecolor{dkred}{rgb}{0.5,0,0}
\definecolor{orange}{rgb}{0.9,0.5,0.3}
\definecolor{violet}{rgb}{0.7,0,0.7}
\usepackage{listings,lstcoq}

\usepackage{quiver}
\usepackage{xspace}
\usetikzlibrary{shapes.misc}

\newcommand\C{\mathbf{C}}
\newcommand\D{\mathbf{D}}
\newcommand\op{{op}}
\newcommand\Types{\ensuremath{\mathbf{Types}}\xspace}
\newcommand\finTypes{\ensuremath{\mathbf{finTypes}}\xspace}
\newcommand\Setoids{\ensuremath{\mathbf{Setoids}}\xspace}
\newcommand\SGraphs{\ensuremath{\mathbf{SGraphs}}\xspace}
\newcommand\MGraphs{\ensuremath{\mathbf{MGraphs}}\xspace}
\newcommand\Type{\lstinline{Type}\xspace}
\newcommand\Prop{\lstinline{Prop}\xspace}
\newcommand\Morphism{\lstinline{Morphism}\xspace}

\setcopyright{cc}
\setcctype{by}
\acmDOI{10.1145/3779031.3779105}
\acmYear{2026}
\copyrightyear{2026}
\acmISBN{979-8-4007-2341-4/2026/01}
\acmConference[CPP '26]{Proceedings of the 15th ACM SIGPLAN International Conference on Certified Programs and Proofs}{January 12--13, 2026}{Rennes, France}
\acmBooktitle{Proceedings of the 15th ACM SIGPLAN International Conference on Certified Programs and Proofs (CPP '26), January 12--13, 2026, Rennes, France}
\received{2025-09-12}
\received[accepted]{2025-11-13}

\begin{document}

\lstset{language=Coq}
\newtheorem{remark}[theorem]{Remark}

\title[Adhesive Category Theory for Graph Rewriting in Rocq]%
{Adhesive Category Theory\\for Graph Rewriting\\in Rocq}

\author{Samuel Arsac}
\email{samuel.arsac@ens-lyon.fr}
\orcid{0009-0007-6508-7636}
\affiliation{
  \institution{ENS de Lyon, CNRS, UCBL1, LIP, Plume team, UMR 5668}
  \city{Lyon}
  \country{France}
}
\author{Russ Harmer}
\email{russell.harmer@ens-lyon.fr}
\orcid{0000-0002-0817-1029}
\affiliation{
  \institution{CNRS, ENS de Lyon, UCBL1, LIP, Plume team, UMR 5668}
  \city{Lyon}
  \country{France}
}
\author{Damien Pous}
\email{damien.pous@ens-lyon.fr}
\orcid{0000-0002-1220-4399}
\affiliation{
  \institution{CNRS, ENS de Lyon, UCBL1, LIP, Plume team, UMR 5668}
  \city{Lyon}
  \country{France}
}

\acmCodeLink{https://gitlab.com/SamuelArsac/graph-rewriting}

\begin{abstract}
  We design a Rocq library about adhesive categories, using Hierarchy Builder (HB).
  It is built around two hierarchies. The first is for categories, with usual categories at the bottom and adhesive categories at the top, with weaker variants of adhesive categories in between. The second is for morphisms (notably isomorphisms, monomorphisms and regular monomorphisms). Each level of these hierarchies is equipped with several interfaces to define instances.
  
  We cover basic categorical concepts such as pullbacks and equalizers, as well as results specific to adhesive categories. Using this library, we formalize two central theorems of categorical graph rewriting theory: the Church-Rosser theorem and the concurrency theorem.
  We provide several instances, including the category of types, the category of finite types, the category of simple graphs and categories of presheaves.
  We detail the implementation choices we made and report on the usage of HB for this formalization work.
\end{abstract}


\maketitle

\section{Introduction}

Categorical graph rewriting, or graph transformation~\cite{rozenbergHandbookGraphGrammars1997,monograph}, is a framework for defining rewriting rules on graphs, together with the double-pushout (DPO) operational semantics, using categories to abstract away the differences between specific concrete graph models. It has found various applications in computer science, such as model-based software engineering~\cite{heckelGraphTransformationSoftware2020} and in defining the syntax and semantics of visual and object-oriented specification languages~\cite{handbook2}; in mathematical physics, such as the rewriting of string diagrams~\cite{bonchiStringDiagramRewrite2022}; and in modelling (bio-)chemical systems, where rewriting rules model the interactions between particles or molecules~\cite{danosConstrainingRulebasedDynamics2013,andersenSoftwarePackageChemically2016}.

In this domain, one often uses \emph{adhesive} categories~\cite{lackAdhesiveQuasiadhesiveCategories2005,garnerAxiomsAdhesiveQuasiadhesive2012}; these include all presheaf categories, hence many categories of multigraphs. However, there are important categories of graphs that are not adhesive. This motivated the introduction of weaker notions to cover those cases and still make it possible to establish good properties of the DPO semantics. For example, the category of simple graphs is only \emph{rm-quasiadhesive}~\cite{johnstoneQuasitoposesQuasiadhesiveCategories2007,garnerAxiomsAdhesiveQuasiadhesive2012} while the category of term graphs is \emph{rm-adhesive} (also called \emph{quasiadhesive} in the earlier literature)~\cite{corradiniTermGraphsAdhesive2005}.

The definitions of these classes of categories involve standard notions such as pullbacks, pushouts and monomorphisms, and each class admits several equivalent definitions. Proving the various implications existing between these definitions is sometimes non-trivial, and their formalization can be seen as an interesting test case for the development of general purpose category theory libraries.
Even more importantly, and as usual for formalization, a difficult task consists in structuring the definitions so that lemmata can be stated and used in their most general form, and instances can be defined in the most efficient way and inferred automatically. This is the main challenge of the present work.

To this end, we choose to use Hierarchy Builder (HB)~\cite{cohenHierarchyBuilderAlgebraic2020}, which provides a high-level language to define hierarchies of structures, with advanced features such as multiple inheritance. HB checks the consistency of these high-level descriptions and translates them into appropriate Rocq records, canonical structures and coercion declarations~\cite{garillotPackagingMathematicalStructures2009}, to obtain concise notations and robust inference mechanisms without further user input.

Our formalization currently includes (i) a general purpose category theory library; (ii) the definitions of adhesive, rm-adhesive and rm-quasiadhesive categories; (iii) two important theorems providing several characterizations of those structures; (iv) two key theorems from categorical graph rewriting; and (v) prototypical instances (types, setoids, finite types, simple graphs, presheaves and slice categories).

The abstract library is axiom-free, and we detail the axioms which are needed for some of the concrete instances.

The development is available on Gitlab\footnote{\url{https://gitlab.com/SamuelArsac/graph-rewriting}} and as an archive~\cite{arsacArtifactAdhesiveCategory2025}, %
with a \textsc{README} file linking the theorems in this paper to the relevant parts of the code.

\paragraph*{Outline.}

In Section~\ref{sec:general}, we detail our implementation of the basic category theory\footnote{While we recall definitions for concepts specific to adhesive category theory, we assume familiarity with the basic concepts of category theory~\cite{pierce:basiccat}.}
that we require and its organization using HB. A specificity of our work is that we define not only a hierarchy of categories, but also a hierarchy of morphisms (isomorphisms, monomorphisms, epimorphisms and their regular and split variants). This is important, as these concepts are pervasive and we need good inference mechanisms to work with them efficiently. We review the notations we set up, how we cover bundled and unbundled definitions, and how we exploit categorical duality.

In Section~\ref{sec:specific}, we move to the specifics of categories for graph transformation. We recall the key definitions and theorems, explain how to extend the previous standard hierarchies with these notions, in a modular fashion, while conforming to the forgetful inheritance policy of HB, and give an overview of the main theorems we have formalized for (rm-(quasi))adhesive categories.
In Section~\ref{sec:rewriting} we briefly present categorical graph rewriting and the two theorems from it we have formalized. We discuss the formalization of concrete instances in Section~\ref{sec:instances}.
We conclude with a discussion of the results (Section~\ref{sec:ccl}), related work (Section~\ref{sc:rw}) and directions for future work (Section~\ref{sc:fw}).

\section{General Category Theory}
\label{sec:general}

\subsection{Equality on Morphisms}
\label{ssec:intro:equality}

The first question when formalizing categories in type theory is ``how to compare morphisms?''.
A simple approach uses propositional equality, and rests on axioms such as functional extensionality and proof irrelevance (e.g., to compare natural transformations) and some forms of choice and predicate extensionality (e.g., to build quotient types).

To reduce the use of axioms while keeping propositional equality, one may consider univalent type theory and formalize univalent categories~\cite{ahrensUnivalentCategoriesRezk2015,voevodskyUniMathComputercheckedLibrary,agda-unimath}.
However, we still need functional extensionality early on for several constructions, the univalence axiom to define the category of (h)sets and functions, and Rezk completion~\cite{ahrensUnivalentCategoriesRezk2015} to deal with non-univalent categories. 

To avoid axioms completely, we can instead formalize $\mathcal E$-categories~\cite{DubricDybjierScott98:pcategories,Kinoshita97:ecategories}, where each homset is equipped with a user-defined equivalence relation\footnote{Such relations are usually defined uniformly over all pairs of objects, although this is not a formal requirement.}, and equations between morphisms hold only up to these equivalence relations. Like in~\cite{wiegleyCategoryTheoryCoq2014,huFormalizingCategoryTheory2021}, this is what we do here, using Rocq support for \emph{setoid rewriting}~\cite{Sozeau09}.
The main advantage is that we keep all the abstract parts of the library axiom-free, postponing the potential use of axioms to specific instances. In particular, we obtain that the category of setoids and setoid morphisms is rm-quasiadhesive, and thus that the graph transformation theorems we formalized hold in that category without any axioms.
We further discuss this choice in Section~\ref{ssec:discussion:setoids}.

\subsection{The Hierarchies of Categories and Morphisms}
\label{ssec:hierarchies}

We start with the definition of categories from the examples found in the repository of HB (adapted for $\mathcal E$-categories, which we simply call categories in the sequel).
It is defined in three steps:
\begin{itemize}
\item quivers: objects and homsets;
\item precategories: quivers with identity arrows and composition;
\item categories: precategories with equations for identities and associativity of composition.
\end{itemize}

While our development does not strictly speaking require such a level of detail in the definition of categories, this is a good example of the way we use a hierarchy: we define structures that inherit from each other (e.g., every category is a quiver) and  define operations and lemmata at the lowest (i.e., most general) point at which they can be stated or proved. For example, the notion of terminal object can be defined in any quiver; the notion of isomorphism and the fact that terminal objects are unique up to unique isomorphism require precategories; and the fact that the composition of two isomorphisms is an isomorphism holds only in categories.

Alongside this hierarchy of categories, we define a hierarchy of morphisms: in addition to monomorphisms, epimorphisms, and isomorphisms, we also need the concepts of \emph{regular} and \emph{split} monomorphisms, as well as their duals: regular and split epimorphisms. (A regular monomorphism is a morphism which is the equalizer of \emph{some} pair of parallel morphisms; a split monomorphism is a morphism with a retraction.) In the rest of this paper, we abbreviate `epimorphism' and `monomorphism' as `epi' and `mono' respectively, as is common in the literature.

\begin{figure}
  \centering
  \quad
  \begin{tikzcd}
    {\texttt{Cat}} \\
    {\texttt{PreCat}} \\
    {\texttt{Quiver}}\\
    {\text\Type}
    \arrow[from=1-1, to=2-1]
    \arrow[from=2-1, to=3-1]
    \arrow[from=3-1, to=4-1]
  \end{tikzcd}
  \hfill
  \begin{tikzcd}[column sep=.5em]
    & {\texttt{Iso}} \\
    {\texttt{SplitEpi}} && {\texttt{SplitMono}} \\
    {\texttt{RegEpi}} && {\texttt{RegMono}} \\
    {\texttt{Epi}} && {\texttt{Mono}} \\
    & {\text\Morphism}
    \arrow[from=1-2, to=2-1]
    \arrow[from=1-2, to=2-3]
    \arrow[from=2-1, to=3-1]
    \arrow[from=2-3, to=3-3]
    \arrow[from=3-1, to=4-1]
    \arrow[from=3-3, to=4-3]
    \arrow[from=4-1, to=5-2]
    \arrow[from=4-3, to=5-2]
  \end{tikzcd}   
  \quad
  \caption{The two hierarchies in the library. An arrow from a structure A to a structure B means that every instance of A is also an instance of B, i.e., A \emph{inherits} from B.}
  \label{fig:basic:hierarchies}
\end{figure}
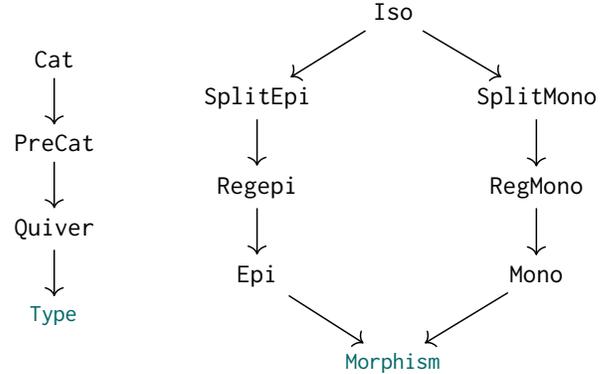

These two hierarchies are depicted in Figure~\ref{fig:basic:hierarchies}. We will extend them with more specialized notions, when moving to adhesive category theory, and it is crucial to have proper tool support for this: our use case here, of specializing to adhesive categories, is just one instance of a frequent activity in the context of formalizing large hierarchies of structures. More specifically, we need to be able to:
\begin{enumerate}
\item use freely any definition, notation or lemma available, for a given structure, in all structures that inherit from it (i.e., all structures above it in the hierarchy);
\item add new structures without breaking existing code, possibly in the middle of the hierarchy;
\item provide instances incrementally and by various means.
\end{enumerate}

In Rocq, there are two principal ways of implementing such hierarchies: typeclasses~\cite{SPITTERS_VAN_DER_WEEGEN_2011,wiegleyCategoryTheoryCoq2014} and canonical structures~\cite{mahboubi:hal-00816703}. Typeclasses are often easier to set up at first, but are rather fragile and may become inefficient in large hierarchies~\cite[Section~12]{Baanen25}. Canonical structures have been used at a large scale in mathcomp~\cite{mathcomp:book}, and have been proved robust and efficient, but require expert knowledge of the unification algorithms.
Some developers have used a mixture of canonical structures and typeclasses~\cite{pousKleeneAlgebraTests2013}.
HB provides an abstract layer and implements packed classes~\cite{garillotPackagingMathematicalStructures2009}, using canonical structures, on behalf of the end user. We have chosen to experiment with its capabilities and we report on its current limitations.

Inference in packed classes is driven by a single \emph{index}. In our hierarchy of categories, the index is the type of objects. For example, we register the category of functors and natural transformations with the type of functors, so that this category is inferred automatically when we look for morphisms between functors.
HB ensures that such inferences are canonical: there is only one way to infer a given structure for a given index, which is crucial for robustness. In cases where we need to work with different structures sharing the same index, we can use \emph{aliases}. 
For instance, we define an alias \Types for \Type, and we register the category of types and functions with \Types. While \Types unfolds to \Type, the latter category is inferred only when using \Types. This makes it possible to register other categories whose objects are types, using other aliases.

In our hierarchy of morphisms, the index for canonical inference is the morphism itself.
Once we have shown that the identity morphism is an isomorphism, we register this information, so that it gets inferred automatically. Similarly, we register the fact that the composition of two isomorphisms is an isomorphism, so that composite isomorphisms can be inferred bottom-up, driven by the composition symbol. Note that it is not possible to declare alternative ways for a composition to be an isomorphism, as this would break canonicity; still, we may use aliases to manually guide inference if needed.

The most important concept for packed classes is that of \emph{mixin}: a datatype depending on the index and on other mixins on that index. For example:
\begin{itemize}
\item the mixin for quivers has no dependency but the index type of objects, and consists of a setoid (a type with an equivalence relation) for each pair of objects; 
\item the mixin for precategories depends on the mixin for quivers and consists of identity arrows and composition; 
\item the mixin for categories depends on the mixin for precategories and consists of proofs that the identities and composition satisfy the expected equations;
\item mixins for morphisms define individual properties of morphisms and so never depend on each other, only on their index, i.e., a morphism;
\item the mixin for monos consists of the proof that the morphism is a mono---a property in \Prop;
\item the mixin for a regular mono consists of the pair of morphisms of which it is an equalizer and the corresponding proof; similarly, the mixin for a split mono consists of a retraction together with the corresponding proof. In both cases this additional information must be in \Type since it is computationally relevant (see Remark~\ref{rmk:compcontent:setoids} below for a full discussion).
\end{itemize}

\emph{Structures} in HB are then defined as collections of mixins, and \emph{inheritance} is defined by inclusion: a structure A inherits from a structure B if it contains \emph{all} its mixins.
This point is very important as it means that HB only implements \emph{forgetful} inheritance~\cite{cohenHierarchyBuilderAlgebraic2020}. This policy has the advantage of robustness---but has several consequences for the way we must structure the code (discussed at the ends of sections~\ref{ssec:vk} and~\ref{subsub:adh}).

In our development, structures are always defined with exactly one mixin in addition to those of the structures it inherits from. We have mentioned some of these mixins above.
Observe that, with these definitions, a regular mono contains two mixins: that for regular monos and that for monos, even though the former implies the latter. If we remove the latter for the sake of minimality, the inheritance from regular monos to monos would be \emph{non-forgetful}. For the same reason, an isomorphism consists of seven mixins (cf. Figure~\ref{fig:basic:hierarchies}). 

To cope with these redundancies, we use the mechanism of \emph{builders} provided by HB: we prove and declare that the mixin for monos can be derived from that for regular monos, so that it suffices to provide the latter in order to build concrete instances. (See Section~\ref{ssec:regularmonos} for details.)
We proceed similarly with other standard implications (e.g., split mono implies regular mono), but this is still not enough. For example, it is well known that a regular mono which is also an epi is an isomorphism; and dually, every regular epi which is also a mono is an isomorphism. Thus we have at least two other ways to construct an isomorphism than to show both split epi and split mono, as inheritance would suggest. To this end, we use \emph{factories}, which are  mixins that do not appear in structures, but can be used to build other mixins. For the previous example, we declare an empty factory, depending on the two mixins for regular monos and epis, and we define builders constructing all other mixins from this data (and dually for the second alternative).

\subsection{Notations}
\label{ssec:notations}

Clear and concise notations with overloading are crucial in formalized mathematics~\cite{GonthierAABCGRMOBPRSTT13}.
Combined with implicit arguments and Rocq's notation system, the mixture of canonical structures and (reversible) coercions used by HB makes it possible to obtain a notation system which remains close to textbook notations.

For example, we just use \lstinline{A~>B} for the type of arrows from \lstinline{A} to \lstinline{B} in a category.
Since categories are indexed by their object type, the type of \lstinline{A} and \lstinline{B} canonically determines that category.
If they are spans, the arrow denotes a span morphism; if they are functors, the arrow denotes a natural transformation, etc. Similarly, we use the symbol \lstinline{idmap} for identities, $(\circ)$ for composition, $(\equiv)$ for morphism equality, \mbox{$(\simeq)$} for isomorphisms, whatever the category. For situations where we want the notation to enforce the choice of category, and thus the type of objects, we also set up variants indexed by the category (e.g., \lstinline{A~>$_\C\,$B}).

A recurrent pattern in category theory is the existence of a unique morphism satisfying a given property, where unicity is up to morphism equivalence.
As in Wiegley's library~\cite{wiegleyCategoryTheoryCoq2014}, we use a single notation (\lstinline{sigma!}\!\!) and a few tools to manipulate such statements.
We define this notation in \Type as the witness is most often computationally relevant (e.g., we need full access to a mediating arrow in a pullback, in order to be able to use it to construct new morphisms).

\subsection{Bundling}
\label{ssec:bundling}

Structures defined with HB are \emph{bundled}. For example, a category is a single package containing the type of objects, the type of the homsets, the identities, etc. This is convenient when we want to quantify on structures, and yields concise and readable proof contexts: we may have a single variable \lstinline{i: A $\simeq$ B} that denotes an isomorphism, rather than a variable for the underlying morphism and a separate assumption about it actually being an isomorphism.

However, there are situations where we need unbundled structures.
Consider for instance the statement: ``pullbacks along isomorphisms are isomorphisms''.
The isomorphism taken as input can be fully bundled, but the output one cannot: it is the purpose of the lemma to promote it from  being a mere morphism to being an isomorphism; so we need a name for the predicate ``being an isomorphism''.

This tension between bundled and unbundled definitions also occurs with concepts outside our current hierarchies.
Consider for instance pullbacks, for which we may use several definitions, each with its own pros and cons:
\begin{enumerate}
\item Being a pullback can be stated as a property of four morphisms organized in a square:
\begin{lstlisting}
Definition isPullback {\C: PreCat} {A B C D: \C}
  (f: A~>C) (g: B~>C) (p: D~>A) (q: D~>B) := 
  f°p ≡ g°q /\
  forall E (p': E~>A) (q': E~>B),
  sigma!u: E~>D, p' ≡ p°u /\ q' ≡ q°u.
\end{lstlisting}
The corresponding diagram is recalled below:
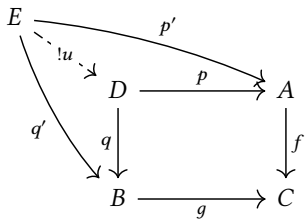
\begin{center}
  \begin{tikzcd}[row sep=scriptsize]
    E \\
    & D && A \\
    \\
    & B && C
    \arrow["!u"{description}, dashed, from=1-1, to=2-2]
    \arrow["{p'}", curve={height=-6pt}, from=1-1, to=2-4]
    \arrow["{q'}"', curve={height=6pt}, from=1-1, to=4-2]
    \arrow["{p}", from=2-2, to=2-4]
    \arrow["{q}"', from=2-2, to=4-2]
    \arrow["f", from=2-4, to=4-4]
    \arrow["g"', from=4-2, to=4-4]
  \end{tikzcd}
\end{center}
This fully unbundled definition is well-suited to state and prove standard facts, such as the pasting lemma of pullback squares, stability of monos under pullbacks or lemmata about isomorphisms that can be moved around the corners of a pullback square: 
\begin{lstlisting}
Lemma isMono_stable_pb ...:
  isPullback f g p q -> isMono f -> isMono p.
Lemma isPullback_iso_A ...(i: A\_≃A')(f': A'~>C): 
  isPullback f' g (i°p) q
  <-> isPullback (f'°i) g p q.
\end{lstlisting}
(In the first lemma, \lstinline{isMono} is our \emph{unbundled} predicate for ``being a mono''.)
\item However, the above definition is not convenient to define categories with all pullbacks. Indeed, in such a case, it is preferable to have the (bundled) type of spans that complete a given cospan into a pullback:
\begin{lstlisting}
Definition Pullback {\C: PreCat} {A B C: \C}
  (f: A~>C) (g: B~>C) :=
  sigma D (p: D~>A) (q: D~>B), isPullback f g p q.
\end{lstlisting}
This type is not convenient for stating the aforementioned lemmata, but to require all pullbacks in a category $\C$, we can simply ask for a function 
\begin{lstlisting}
pullback: forall {A B C: \C} (f: A~>C) (g: B~>C),
  Pullback f g
\end{lstlisting}
\item When it comes to proving that pullbacks are unique up to unique isomorphisms, the best definition is the most categorically-minded one: a pullback is a terminal object in the category of spans completing the given cospan into a commuting square. We prove this characterization: it requires us to define subcategories and the category of spans, but uniqueness then comes for free as a special case of unicity of terminal objects. 
\end{enumerate}
In the end, we keep both a fully unbundled definition (stating that a square of morphisms \emph{is} a pullback) and a partly bundled one (stating that a cospan \emph{has} a pullback), and we set up appropriate lemmata to move back and forth between them. We proceed similarly with all other concepts in the library. In particular, the aforementioned lemma about stability of monos along pullbacks is rather stated as follows, where \lstinline{A$\hookrightarrow$C} is our notation for the type of monos from \lstinline{A} to \lstinline{C}. 
\begin{lstlisting}
Lemma Mono_stable_pb {\C: Cat} {A B C D: \C}
  (f: A$\hookrightarrow$C) (g: B~>C) (p: D~>A) (q: D~>B):
  isPullback f g p q -> isMono p.
\end{lstlisting}

\subsection{A Concrete Example: Regular Monos}
\label{ssec:regularmonos}

To give a concrete example of how things are set up with HB, here is how we define regular monos in the library.

First, recall that an \emph{equalizer} of a pair of parallel morphisms $f$ and $g$ is the most general morphism $e$ such that $f \circ e = g \circ e$, as illustrated in the diagram below.
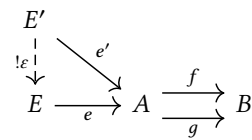
\begin{center}
\begin{tikzcd}
	{E'} \\
	E & A & B
	\arrow["{!\varepsilon}"', dashed, from=1-1, to=2-1]
	\arrow["{e'}", from=1-1, to=2-2]
	\arrow["e"', from=2-1, to=2-2]
	\arrow["f", shift left=2, from=2-2, to=2-3]
	\arrow["g"', shift right=2, from=2-2, to=2-3]
\end{tikzcd}
\end{center}

In Rocq we define the following predicate:
\begin{lstlisting}
Definition isEqualizer {\C: PreCat} {A B E: \C}
  (f g: A~>B) (e: E~>A) := f°e ≡ g°e /\
  forall E' (e': E'~>A), f°e' ≡ g°e' ->
  sigma! $\varepsilon$: E'~>E, e' ≡ e°$\varepsilon$.
\end{lstlisting}
A regular mono is then simply a morphism that is an equalizer for \emph{some} pair of morphisms, and we can prove that every regular mono is indeed a mono:
\begin{lstlisting}
Definition isRegMono {\C E A} (m: E~>$_\C\,$A) :=
  sigma B (f g: A~>B), isEqualizer f g m.
Lemma isRegMono_isMono {\C E A} (m: E~>$_\C\,$A):
  isRegMono m -> isMono m.
\end{lstlisting}
We can now register the structure in HB as follows:
\begin{lstlisting}
HB.mixin Record IsRegMono \C\_E A (m: E~>$_\C\,$A) :=
  {rm_prop: isRegMono m}.  
HB.structure Definition RegMono \C\_E A :=
  {m of Mono \C\_E A m & IsRegMono \C\_E A m}.
HB.builders Context \C\_E A m of IsRegMono \C\_E A m.
  HB.instance Definition _ :=
    IsMono.Build m (isRegMono_isMono m rm_prop).
HB.end.
\end{lstlisting}
First, we define a new mixin; then we define a new structure \lstinline{RegMono}, consisting of this new mixin together with all the mixins (in fact just one) of the structure \lstinline{Mono}; and finally, we register the previous lemma as a builder from the new mixin to the mixin of \lstinline{Mono}. This builder makes it possible to define regular monos by providing \emph{only} the expected property, while the redundant definition of the structure \lstinline{RegMono} ensures \emph{forgetful} inheritance from regular monos to monos.

From this point, we can use both the type \lstinline{RegMono A B} of regular monos from \lstinline{A} to \lstinline{B} and the unbundled predicate \lstinline{isRegMono f} characterizing such morphisms.

\begin{remark}
  \label{rmk:compcontent:setoids}
  Like for pullbacks, ``$e$ being an equalizer of $f$ and $g$'' cannot be formalized in \Prop. The first requirement that $f \circ e = g \circ e$ is indeed a proposition, but the second one, the fact that we get a mediating morphism $\epsilon$ from any other morphism $e'$ satisfying the same property, is not. Indeed, this requirement is essentially a function from morphisms to morphisms, and we must be able to use this function freely in various constructions---not just in propositional contexts.

  For the same reason, ``being a regular mono'' carries computational content, and should be formalized in \Type rather than in \Prop.
  This is why we use sigma types rather than existential quantifications (cf. Section~\ref{ssec:notations}).
  
  Now, a given mono has many ways of being regular, but unlike pullbacks which are unique up to unique isomorphism, those ways of being regular need not be equivalent. Still, we do not want to
  distinguish regular monos based on this additional computational content. This is straightforward to achieve with our setoid-based development: we can choose to compare regular monos via their underlying morphism. In other settings, including proof-irrelevant or univalent ones, we would typically need to take a proper quotient.

  In the same vein, a given mono may have many retractions, thus many
  inequivalent ways of being split. While we need full access (i.e.,\ in \Type) to the retraction of a split mono, we do not want to distinguish split monos
  that differ only by their choice of retraction. Here again, the setoid-based approach makes it possible to avoid proper quotients and related axioms.
\end{remark}

\subsection{Duality}
\label{ssec:duality}

Another important point in category theory is duality: every abstract concept or lemma has a dual: epis/monos, equalizers/coequalizers, initial/terminal objects and their unicity, pushouts/pullbacks and their pasting lemmata, etc.
Like in other works~\cite{pousKleeneAlgebraTests2013,ahrensUnivalentCategoriesRezk2015,wiegleyCategoryTheoryCoq2014,huFormalizingCategoryTheory2021,leanmathlib}, we try to exploit duality as much as possible, in order to avoid code duplication.

The central tool is the definition of the dual $S^\op$ of a structure $S$ (be it a quiver, precategory, category or one of the various kinds of morphism): by applying a lemma on a generic structure $S$ to the structure $S^\op$, we get the dual statement.

Since the structures $S$ and $S^\op$ share the same type of objects, we need to disambiguate canonical structure inference. To this end, we simply use an alias: we define an identity function \lstinline{catop: Type->Type}, which we bind to the notation \lstinline{S^op} and use to annotate object types for which we want the dual structures to be inferred. For morphisms, we use the same technique, using the following identity function:
\begin{lstlisting}
Definition morphop {\C: Quiver} {A B}
  (f: A ~>$_\C$ B): B $\rightsquigarrow_{\C^\op}$ A := f.    
\end{lstlisting}
Then we systematically declare instances for dual structures; for example, we write 
\begin{lstlisting}
HB.instance Definition _ {\C} (A B: \C) (m: Mono A B)
  := IsEpi.Build \C^op\_B A (morphop m) (mono_prop m).  
HB.instance Definition _ {\C} (A B: \C) (m: Epi A B)
  := IsMono.Build \C^op\_B A (morphop m) (epi_prop m).  
\end{lstlisting}
so that \lstinline{morphop m} is automatically recognized as an epi whenever \lstinline{m} is a mono, and vice-versa. 

Using this strategy, we manage to reuse proofs for most dual statements. For example,
\begin{lstlisting}
Lemma Epi_stable_po {\C} ... (f: C$\twoheadrightarrow$A)...:
  isPushout f g p q -> Epi p.
Proof. exact: (Mono_stable_pb (\C:=\C^op)). Qed
Lemma isPushout_iso_A {\C} ...A' (i: A'≃A)(f': C~>A'): 
  isPushout f' g (p°i) q <-> isPushout (i°f') g p q.
Proof. exact: (isPullback_iso_A (iso_op i)). Qed.
\end{lstlisting}
For these two proofs to typecheck, we need two things.
\begin{enumerate}
\item The types \lstinline{@isPushout $\C$} and \lstinline{@isPullback $\C^\op$} should be definitionally equal. Moreover, for readability and end-user experience, neither the definition of pushout nor that of pullback should involve $\C^\op$: we want duplicated explicit definitions. The combination of these two requirements forces us to use generic sigma types rather than dedicated records for such definitions: in Rocq, two record types are definitionally equal iff they share the same name and have definitionally equal parameters. 
\item We need to explicitly cast isomorphisms in $\C$ to isomorphisms in $\C^\op$ (the \lstinline{iso_op} function in the second proof). Indeed, these types are only equivalent, even up to reversing arguments, even with proof irrelevance: if $A\simeq_\C B$ is a record with three fields:
\begin{lstlisting}
{i: A~>$_\C\,$B; j: B~>$_\C\,$A; _: i°j ≡ id /\ j°i ≡ id}
\end{lstlisting}
then $B\simeq_{\C^\op} A$ is the record:
\begin{lstlisting}
{i: A~>$_\C\,$B; j: B~>$_\C\,$A; _: j°i ≡ id /\ i°j ≡ id}
\end{lstlisting}
(Note that the two conjuncts have been swapped, so that we may at most prove the two types propositionally equal, using propositional extensionality.)
\end{enumerate}

Another natural desideratum is that the dual operation should be as involutive as possible: $(S^\op)^\op=S$. We would typically need this equality, definitionally, to dualize the proofs of statements mentioning both a category and its dual.
However:
\begin{itemize}
\item since \Prop is proof-relevant in Rocq, we need to ensure that dual is involutive even on proofs: this is immediate for quivers and precategories, since they do not include proofs, but for categories we should use the standard trick of duplicating the associativity proof in the mixin defining the structure~\cite{wiegleyCategoryTheoryCoq2014,voevodskyUniMathComputercheckedLibrary,huFormalizingCategoryTheory2021}, and turn the textbook mixin into a factory. Such a technique may become heavy for richer structures, however (e.g., bicategories).
\item HB structures and mixins are defined as records, so that we need the eta law on records to get definitional equality when the structure $S$ is abstract. Rocq supports this for \emph{primitive} records, but having eta for HB structures may defeat the canonical structure inference mechanism\footnote{https://github.com/math-comp/math-comp/pull/462\#issuecomment-598130155}.
\end{itemize}
We did not need such techniques in the present development.
However, in the long term, we would prefer to develop support for transporting results along equivalences of categories, which should make it possible to avoid the aforementioned contortions and rely on robust mathematics: $(\C^\op)^\op$ is equivalent to $\C$, the category of isomorphisms in $\C^\op$ is equivalent to that on $\C$, etc.

\section{Adhesive Category Theory}
\label{sec:specific}

In the next two sections, we describe the key concepts of adhesive category theory and their application to prove two important theorems in graph transformation in an abstract fashion. In both cases, we aim to provide a brief overview of the mathematical ideas and then present the salient aspects of our formalization---with a focus on the issue of describing this theory in a modular way, based on the generic category theory library of the previous section.

\subsection{Van Kampen Squares and Adhesive Morphisms}
\label{ssec:vk}

We first define three properties of pushout squares and two new classes of morphisms.
We provide these definitions to give an idea of the concepts that we need to manipulate; they can be taken as black boxes in what follows.

\begin{definition}[Van Kampen square~\cite{lackAdhesiveQuasiadhesiveCategories2005}]
  A pushout is
  \begin{itemize}
  \item \emph{stable} iff, for every commutative cube with the arrows as in the diagram below, where the bottom face is the pushout square and where the vertical faces are pullbacks, the top face is a pushout;
  \item \emph{exact} iff, for every commutative cube with the arrows as in the diagram below, where the bottom face is the pushout square, the left and back faces are pullbacks, and the top face is a pushout, the front and right faces are pullbacks;
  \item \emph{Van Kampen} if it is both stable and exact.
  \end{itemize}
\end{definition}
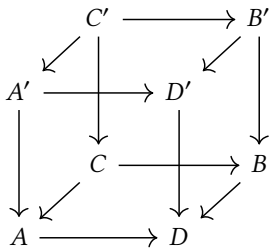
\begin{center}
\begin{tikzcd}[column sep=small,row sep=scriptsize]
	& {C'} && {B'} \\
	{A'} && {D'} \\
	& C && B \\
	A && D
	\arrow[from=1-2, to=1-4]
	\arrow[from=1-2, to=2-1]
	\arrow[from=1-2, to=3-2]
	\arrow[from=1-4, to=2-3]
	\arrow[from=1-4, to=3-4]
	\arrow[from=2-1, to=2-3]
	\arrow[from=2-1, to=4-1]
	\arrow[from=2-3, to=4-3]
	\arrow[from=3-2, to=3-4]
	\arrow[from=3-2, to=4-1]
	\arrow[from=3-4, to=4-3]
	\arrow[from=4-1, to=4-3]
      \end{tikzcd}
\end{center}
The importance of Van Kampen squares for graph transformation was first recognized in~\cite{lackAdhesiveQuasiadhesiveCategories2005} where they were used to abstract recurring reasoning motifs that occur in proofs of the Church-Rosser and concurrency theorems in concrete models.

The two key properties of Van Kampen squares are that they preserve monos and that they are also pullbacks~\cite{lackAdhesiveQuasiadhesiveCategories2005}.

\begin{definition}[Adhesive morphisms~\cite{garnerAxiomsAdhesiveQuasiadhesive2012}]
  A morphism is
  \begin{itemize}
  \item \emph{pre-adhesive} iff it has pushouts along any morphism and these pushouts are stable and also pullbacks;
  \item \emph{adhesive} iff its pullbacks along all morphisms are pre-adhesive.
  \end{itemize}
\end{definition}
Adhesive morphisms are pre-adhesive (by pulling back along the identity); pre-adhesive morphisms are regular monos; and isomorphisms are always adhesive.
These classes were first defined in~\cite{garnerAxiomsAdhesiveQuasiadhesive2012}. They identify those morphisms in a category along which pushouts are well-behaved for the purposes of graph transformation.
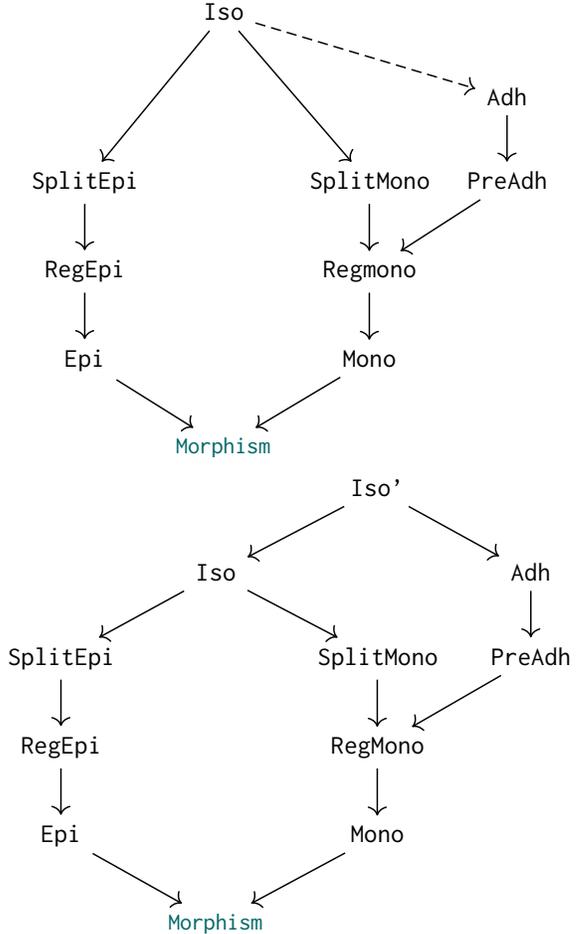
\begin{figure}
  \centering
  \begin{tikzcd}[column sep=tiny]
    & {\texttt{Iso}} \\
    &&& {\texttt{Adh}} \\
    {\texttt{SplitEpi}} && {\texttt{SplitMono}} & {\texttt{PreAdh}} \\
    {\texttt{RegEpi}} && {\texttt{RegMono}} \\
    {\texttt{Epi}} && {\texttt{Mono}} \\
    & {\text\Morphism}
    \arrow[dashed, from=1-2, to=2-4]
    \arrow[from=1-2, to=3-1]
    \arrow[from=1-2, to=3-3]
    \arrow[from=2-4, to=3-4]
    \arrow[from=3-1, to=4-1]
    \arrow[from=3-3, to=4-3]
    \arrow[from=3-4, to=4-3]
    \arrow[from=4-1, to=5-1]
    \arrow[from=4-3, to=5-3]
    \arrow[from=5-1, to=6-2]
    \arrow[from=5-3, to=6-2]
  \end{tikzcd}
  
  \begin{tikzcd}[column sep=small]
    && {\texttt{Iso'}} \\
    & {\texttt{Iso}} && {\texttt{Adh}} \\
    {\texttt{SplitEpi}} && {\texttt{SplitMono}} & {\texttt{PreAdh}} \\
    {\texttt{RegEpi}} && {\texttt{RegMono}} \\
    {\texttt{Epi}} && {\texttt{Mono}} \\
    & {\text\Morphism}
    \arrow[from=1-3, to=2-2]
    \arrow[from=1-3, to=2-4]
    \arrow[from=2-2, to=3-1]
    \arrow[from=2-2, to=3-3]
    \arrow[from=2-4, to=3-4]
    \arrow[from=3-1, to=4-1]
    \arrow[from=3-3, to=4-3]
    \arrow[from=3-4, to=4-3]
    \arrow[from=4-1, to=5-1]
    \arrow[from=4-3, to=5-3]
    \arrow[from=5-1, to=6-2]
    \arrow[from=5-3, to=6-2]
  \end{tikzcd}
  \caption{The extended hierarchy of morphisms: with non-forgetful inheritance (top) or with extended isomorphisms (bottom).}
  \label{fig:extended:morphisms}
\end{figure}

Accordingly, we turn these two classes into HB structures. We can easily add them to our hierarchy above regular monos, with adhesive above pre-adhesive. However, it is not clear what to do for isomorphisms (Figure~\ref{fig:extended:morphisms}). Indeed, declaring adhesive morphisms below isomorphisms would require us to update the definition of isomorphism to include the mixins of adhesive and pre-adhesive morphisms.
While it is possible to reorganize the code in this way, it is not very satisfactory: since adhesive morphisms are rather specific, we wish to be able to treat the previous hierarchy of epis and monos as a library, and adhesive morphisms as an optional module that we can plug in. We are thus left with two options:
\begin{enumerate}
\item Declaring isomorphisms as instances of adhesive morphisms: This gives us the ability to use isomorphisms where adhesive morphisms are expected, but breaks the forgetful inheritance policy of HB. In fact, with such an instance declaration, HB has two ways of equipping an isomorphism with the mixin of, say, a regular mono: either by direct containment, or by using the proofs that it is adhesive, and thus pre-adhesive, and thus a regular mono.
Since these two paths have no particular reason to be equal, this could lead to problematic situations where statements holding along one path cannot be used in contexts mentioning the other. Therefore HB rejects such a definition. 
\item Adding a new isomorphism structure (\lstinline{Iso'}), inheriting from both original isomorphisms and adhesive morphisms. Proceeding in this way, we comply with forgetful inheritance, but we sometimes need to explicitly promote isomorphisms from the core library to isomorphisms from the extended one. This can be done easily using an empty factory, but it nevertheless induces some overhead, and we would have to introduce new copies whenever we need new classes of morphisms below isomorphisms. 
\end{enumerate}
We have chosen the second approach for the time being, hoping that more satisfactory solutions can be designed in the future to deal with this form of inheritance \emph{a posteriori}.

\subsection{Subobjects}

We shall also need the notion of \emph{subobject}, which is a categorical generalization of set-theoretic inclusions. Given an object $X$ in a category $\C$, we take a \emph{subobject} of $X$ to be a mono $m\colon A\hookrightarrow X$.
Subobjects are then the objects of a category, where a morphism from $m\colon A\hookrightarrow X$ to $n\colon B\hookrightarrow X$ is a morphism $f\colon A\to B$ of $\C$ such that $n\circ f=m$.

This category is clearly a preorder. Some authors prefer to take a subobject to be an equivalence class of monos in this preorder, to obtain a partial order. In practice, this ambiguity rarely causes any difficulties; in any case, for our purposes, the simpler notion of subobject as mono suffices, so we prefer to avoid the quotient entirely.

When $\C$ has pullbacks, the category of subobjects has all products, usually called \emph{binary intersections} in this context. Coproducts, which do not always exist, are called \emph{binary unions}. A subobject is \emph{regular} if the underlying mono is.

\subsection{The Hierarchy of Adhesive Categories}

We can now present the three variants of adhesive categories that we have formalized.
Each time, we provide the textbook definition---together with alternative characterizations, if any---and explain the choices we make.

Historically, adhesive categories were first defined in 2004 and subsequently generalized to rm-adhesive~\cite{lackAdhesiveQuasiadhesiveCategories2005} and finally rm-quasiadhesive~\cite{garnerAxiomsAdhesiveQuasiadhesive2012} categories in order to capture larger classes of concrete models. Much of the recent graph transformation literature states theorems in terms of adhesive categories, but in some cases these results actually hold in these more general settings---as we will see in section 4.

The definition of adhesive categories is phrased in more elementary terms than the two weaker variants, as it makes no reference to the concept of regular mono. This allows for a pragmatic approach, avoiding this latter concept, at the cost of restricting the class of concrete models that can be discussed and a dependence on classical axioms to construct certain instances. On the other hand, the use of regular monos in the weaker variants allows for a purer approach encompassing more models and requiring fewer axioms (if any) to construct instances.

\subsubsection{Rm-quasiadhesive Categories}

We start with rm-quasiadhesive categories, the least demanding structure in the hierarchy. The \emph{rm} part of their name stands for regular monos, as they are central to the definition.

\begin{definition}[Rm-quasiadhesive category~\cite{garnerAxiomsAdhesiveQuasiadhesive2012}]
\label{def:rmqadh}
A category is \emph{rm-quasiadhesive} iff
\begin{itemize}
    \item it has all pullbacks,
    \item it has pushouts along regular monos,
    \item pushouts along regular monos are stable, and
    \item pushouts along regular monos are pullbacks.
\end{itemize}
\end{definition}
\noindent
Since having all pullbacks---which means that every cospan has a pullback---is a common requirement, we first define a structure \lstinline{CatPb} for those categories; and, similarly, we define a structure \lstinline{CatRmPo} for categories with pushouts along regular monos---which means that every span where one of the morphisms is a regular mono has a pushout.

We turn the last two requirements of the definition into a new mixin in order to define the structure \lstinline{RmQAdhesive}. We see here---in a slightly more sophisticated case than for quivers, precategories and categories---how stronger structures are built in HB by combining independent pieces of weaker structures.

At this point, our hierarchy contains the following dependencies: 
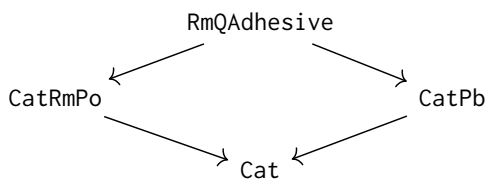
\begin{center}
  \begin{tikzcd}[row sep=scriptsize]
    & {\texttt{RmQAdhesive}} \\
    {\texttt{CatRmPo}} && {\texttt{CatPb}} \\
    & {\texttt{Cat}}
    \arrow[from=1-2, to=2-1]
    \arrow[from=1-2, to=2-3]
    \arrow[from=2-1, to=3-2]
    \arrow[from=2-3, to=3-2]
  \end{tikzcd}
\end{center}

Concerning the theory of rm-quasiadhesive categories, we proved that regular monos are preserved by composition and thus form a subcategory, that pushouts preserve regular monos, and that regular subobjects do have binary unions, although possibly non-regular ones. These results are non-trivial, see~\cite{garnerAxiomsAdhesiveQuasiadhesive2012,lackAdhesiveQuasiadhesiveCategories2005}.

\subsubsection{Rm-adhesive Categories}

Rm-adhesive categories (originally named \emph{quasiadhesive} categories in~\cite{lackAdhesiveQuasiadhesiveCategories2005}) are the next level in the hierarchy of adhesive category theory.

\begin{definition}[Rm-adhesive category~\cite{garnerAxiomsAdhesiveQuasiadhesive2012}, \cite{lackAdhesiveQuasiadhesiveCategories2005}]
\label{def:rmadh}
A category is \emph{rm-adhesive} iff:
\begin{itemize}
    \item it has all pullbacks,
    \item it has pushouts along regular monos, and
    \item pushouts along regular monos are Van Kampen.
\end{itemize}
\end{definition}

Note that the first two properties are the same as for rm-quasiadhesive categories; but the third property appears, at first sight, to be unrelated. However, we have:

\begin{theorem}[{\cite[Thm.~3.3.]{garnerAxiomsAdhesiveQuasiadhesive2012}}]
  \label{thm:rmadhesive}
  In a category $\C$ with all pullbacks, the following are equivalent:
  \begin{enumerate}
  \item $\C$ is rm-adhesive;
  \item regular monos are adhesive, and regular subobjects are closed under binary union;
  \item $\C$ is rm-quasiadhesive, and regular subobjects are closed under binary union.
  \end{enumerate}
\end{theorem}

Since there are three equivalent characterizations, we can choose the most convenient one as the native definition.
While the first point is the textbook definition, the third point in the above theorem shows that rm-adhesive categories actually are rm-quasiadhesive. We thus take this item as the native definition, to optimize modularity: \lstinline{RmAdhesive} only adds the property of ``closure of regular subobjects under binary union'' to \lstinline{RmQAdhesive}.

In HB terms, this means that the second part of point $3$ becomes a mixin, and points 1 and 2 are factories with a dependency on categories with pullbacks, i.e., \lstinline{CatPb}. 
The implications $1\Rightarrow3$ and $2\Rightarrow3$ are proved and declared as builders. This means that they are used by HB as instructions on how to construct the mixins in the hierarchy up to \lstinline{RmAdhesive}, given the data in either factory. This makes it possible to create instances in three different ways, at the user's convenience.

Conversely, the implications $3\Rightarrow1$ and $3\Rightarrow2$ can be seen as being part of the theory of rm-adhesive categories. We state and prove these implications as standard Rocq lemmata; they can be applied whenever we have an instance of a rm-adhesive category:
\begin{itemize}
\item pushouts along regular monos are exact and thus Van Kampen (the remaining part of implication $3\Rightarrow1$);
\item regular monos are adhesive (the remaining part of implication $3\Rightarrow2$, which in fact is a theorem of rm-quasiadhesive categories).
\end{itemize}

\subsubsection{Adhesive Categories}
\label{subsub:adh}

We finally move to adhesive categories, which form the richest structure in our hierarchy: 
\begin{definition}[Adhesive category~\cite{garnerAxiomsAdhesiveQuasiadhesive2012}]
\label{def:adh}
A category is \emph{adhesive} iff:
\begin{itemize}
    \item it has all pullbacks,
    \item it has pushouts along monos, and
    \item pushouts along monos are Van Kampen.
\end{itemize}
\end{definition}

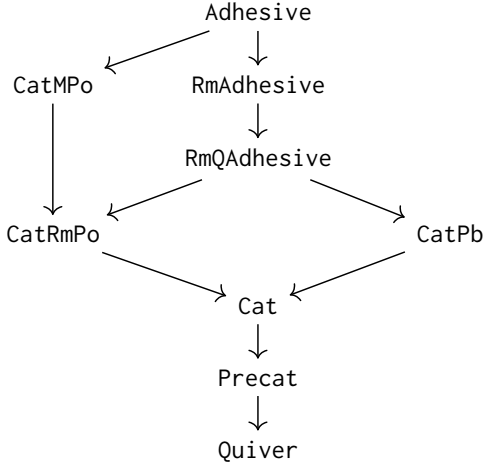
\begin{figure}
  \begin{center}
    \begin{tikzcd}[row sep=scriptsize]
      & {\texttt{Adhesive}} \\
      {\texttt{CatMPo}} & {\texttt{RmAdhesive}} \\
      & {\texttt{RmQAdhesive}} \\
      {\texttt{CatRmPo}} && {\texttt{CatPb}} \\
      & {\texttt{Cat}} \\
      & {\texttt{Precat}} \\
      & {\texttt{Quiver}}
      \arrow[from=1-2, to=2-2]
      \arrow[from=1-2, to=2-1]
      \arrow[from=2-1, to=4-1]
      \arrow[from=2-2, to=3-2]
      \arrow[from=3-2, to=4-1]
      \arrow[from=3-2, to=4-3]
      \arrow[from=4-1, to=5-2]
      \arrow[from=4-3, to=5-2]
      \arrow[from=5-2, to=6-2]
      \arrow[from=6-2, to=7-2]
    \end{tikzcd}
  \end{center}  
  \caption{Extended hierarchy of categories.}
  \label{fig:extended:cats}
\end{figure}

\begin{theorem}[{\cite[Thm.~3.2]{garnerAxiomsAdhesiveQuasiadhesive2012} and \cite[Prop.~6.2]{lackAdhesiveQuasiadhesiveCategories2005}}]
  \label{thm:adhesive}
In a category $\C$ with all pullbacks, the following statements are equivalent:
\begin{enumerate}
    \item $\C$ is adhesive;
    \item all monos are adhesive;
    \item $\C$ has pushouts along monos, those pushouts are stable and are pullbacks;
    \item $\C$ is rm-adhesive and all monos are regular.
    \item $\C$ is rm-quasiadhesive and all monos are regular.
\end{enumerate}
\end{theorem}
(The fifth point is not mentioned in the above references; it follows from the third item of Theorem~\ref{thm:rmadhesive}: unions of regular subobjects exist and are monos in rm-quasiadhesive categories, hence regular monos here.)
We see from this theorem that adhesive categories are rm-adhesive, and that they have pushouts along all monos. The latter requirement is quite common so that we make it into a structure \lstinline{CatMPo}, above \lstinline{CatRmPo}. Then we define adhesive categories as rm-adhesive categories with pushouts along all monos and where all monos are regular.
The hierarchy we obtain is depicted in Figure~\ref{fig:extended:cats}.
We provide four factories to construct adhesive categories, corresponding to items 1,2,3 and~5 in the above theorem (item~4 being superseded by item~5).

As for rm-adhesive categories, we provide the remaining parts of Theorem~\ref{thm:adhesive} as standard Rocq lemmata about adhesive categories.

Note that due to forgetful inheritance, there are two functions providing pushouts in the structure
\lstinline{CatMPo}: those along monos and those along regular monos. While the latter can be deduced from the former, it need not be, and the two functions may return different values on the same input (these values are nevertheless isomorphic, by unicity of pushouts).

To reduce the burden of working explicitly modulo isomorphisms, we carefully write our mixins and statements so that they quantify over all possible pushouts (or other kinds of values in different contexts). For instance the third item in Theorem~\ref{thm:adhesive} is formally modelled as a function providing \emph{some} pushout for each span of a mono and a morphism and the property that \emph{all} pushouts along a mono are stable and are pullbacks---not just the ones provided by the function.

\section{Graph Rewriting Theory}
\label{sec:rewriting}

Adhesive categories and their variants make it possible to establish theorems about graph rewriting systems once and for all, abstracting over the precise choice of graphs (simple, directed, labelled, finite, typed, \dots).
To check whether our library is mature enough to make it possible to formalize state-of-the art results in this field, we have formalized two important theorems~\cite[Theorems~7.7 and~7.10]{lackAdhesiveQuasiadhesiveCategories2005}. For lack of space, we do not provide their precise statements here, only their overall nature and the rather small number of lines we need to formalize them.

\subsection{Elementary Notions of DPO Graph Rewriting}
\label{ssec:elem-rewriting}

\begin{figure}
  \centering
  \begin{tikzpicture}[scale=.7]
    \begin{scope}[every node/.style={rectangle,draw},
      every edge/.style={draw}]
      \node (A) at (0,0) 
          {\begin{tikzpicture}
          \begin{scope}[every node/.style={circle,draw},
            every edge/.style={draw=black}]
              \node[style={fill=black!80}] (A') at (0,0) {};
              \node (B') at (1,0) {};
           \path [-{Latex[length=2mm]}] (A') edge (B');
          \end{scope}
          \end{tikzpicture}};
      \node (B) at (4,0) 
          {\begin{tikzpicture}
          \begin{scope}[every node/.style={circle,draw},
            every edge/.style={draw=black}]
              \node[style={fill=black!80}] (A') at (0,0) {};
              \node (B') at (1,0) {};
          \end{scope}
          \end{tikzpicture}};
      \node (C) at (8,0) 
          {\begin{tikzpicture}
          \begin{scope}[every node/.style={circle,draw},
            every edge/.style={draw=black}]
              \node[style={fill=black!80}] (A') at (0,0) {};
              \node (B') at (1,0) {};
              \node[style={fill=black!40}]  (C') at (0.5,1) {};
           \path [-{Latex[length=2mm]}] (C') edge (B');
          \end{scope}
          \end{tikzpicture}};
    
    \path[-{Latex[length=3mm]}] (B) edge (A);
    \path[-{Latex[length=3mm]}] (B) edge (C);
    \end{scope}
  \end{tikzpicture}
  \caption{An example of a production with simple graphs. The edge is deleted, and a new vertex with an edge is added}
  \label{fig:prod_example}
\end{figure}
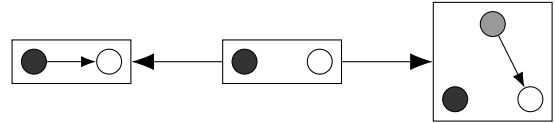

In double-pushout (DPO) graph rewriting, rewriting rules (also called \emph{productions}) consist of a span, the two target objects being the input and the output and the source object being an intermediate state. Such a production is depicted in Figure~\ref{fig:prod_example}. The intuition for the middle object is that it contains the part of the graph (besides the context) that is preserved by the production, i.e.~is neither deleted nor added.

We focus on \emph{linear productions}, in which both morphisms of the span are regular monos.

An application of a rewriting rule is called a \emph{derivation}. Given a production $I \leftarrow K \rightarrow O$ and an object $X$ with a morphism $I \rightarrow X$, a derivation is a diagram of the following shape, where both squares are pushouts.

\[\begin{tikzcd}
	I & K & O \\
	X & Y & Z
	\arrow[from=1-1, to=2-1]
	\arrow[from=1-2, to=1-1]
	\arrow[from=1-2, to=1-3]
	\arrow[from=1-2, to=2-2]
	\arrow[from=1-3, to=2-3]
	\arrow[from=2-2, to=2-1]
	\arrow[from=2-2, to=2-3]
\end{tikzcd}\]
Intuitively, the morphism $I \rightarrow X$ indicates how and where the left-hand side of the production matches the input graph $(X)$, $Z$ is the result of applying the production to $X$, and the morphism $O \rightarrow Z$ shows that the right-hand side of the production appears in that graph.

\subsection{Local Church-Rosser Theorem}
\label{ssec:church-rosser}

The first theorem~\cite[Theorem~7.7]{lackAdhesiveQuasiadhesiveCategories2005} gives a characterization of the local Church-Rosser property from term rewriting systems~\cite{terese}.

It links two properties: \emph{parallel} and \emph{sequential} independence. The first is a property on two derivations starting from the same graph, stating that neither of them deletes the part on which the other production depends. The second is also a property on two derivations, where the second production is applied to the output of the first derivation, stating that the second production neither needs nor deletes what the first production has added to the graph.

The theorem states that the two properties are equivalent. Since parallel independence is symmetric, it is equivalent to sequential independence with the productions in either order. This enables notably the definition of equivalence of derivations in graph grammars and of canonical derivations for each equivalence class~\cite{kreowski77}.

The proof in~\cite{lackAdhesiveQuasiadhesiveCategories2005} is stated for rm-adhesive categories, but it applies to rm-quasiadhesive categories.
The statement and its proof take 2 pages on paper, with diagrams involving up to 11 objects.
Thanks to our library, our formalization took just 80 lines of statements and 78 lines of proof. 

\subsection{Concurrency Theorem}
\label{ssec:concurrency}

The concurrency theorem~\cite[Theorem~7.10]{lackAdhesiveQuasiadhesiveCategories2005} determines when and how we can \emph{compose} two productions into a new one inducing the same derivations as the two productions applied sequentially.
Again this possibility is useful in modelling tools as it enables the grouping of a collection of small rules into a single larger one, e.g., to represent a complex refactoring operation in model-based software engineering~\cite{heckelGraphTransformationSoftware2020} as a single atomic operation.

As above, the proof given in~\cite{lackAdhesiveQuasiadhesiveCategories2005} for rm-adhesive categories carries over to rm-quasiadhesive categories. 
The statement and its proof take 2.5 pages on paper, with diagrams involving up to 19 objects.
We formalized it with 107 lines of statements and 105 lines of proof. 

The fact that our proofs strengthen the statements of these two theorems can be seen as a serendipitous consequence of our formalization effort. However, it also raises the question of the pertinence, or not, of the stronger variants of adhesive categories. In the case of rm-adhesive categories, there is an important theorem---the concurrency theorem for non-linear DPO~\cite{behrConcurrencyTheoremsNonlinear2021}---whose only known proof requires the van Kampen property. A further significant result is the amalgamation theorem whose statement depends on adhesive categories~\cite{golasMultiAmalgamationAdhesiveCategories2010}, but whose proof we have yet to examine in detail in order to formalize (and so it may be that it can be proved in the weaker setting of rm-quasiadhesive categories).

\section{Instances}
\label{sec:instances}

Now we discuss the key instances we have formalized: types, setoids, finite types, simple graphs, slice and functor categories.
In each case, we discuss the required type-theoretic axioms, if any.

\subsection{Types and Functions}
\label{ssec:types}

We have proven that the category \Types with \Type as the type of objects and extensional functions as morphisms\footnote{More precisely, morphisms are functions, but we compare them pointwise.} is adhesive. This is the counterpart in type theory of the category of sets and functions in set theory.
We use the third item of Theorem~\ref{thm:adhesive} to construct this adhesive category.
Proving that it has all pullbacks is easy, but constructing and analysing pushouts is more difficult.

The pushout of two morphisms $f: C \rightarrow A$ and $g: C \rightarrow B$ can be computed by first taking the disjoint union of $A$ and $B$, then quotienting it by the smallest equivalence relation $R$ such that $(f\: c)\: R\: (g\: c)$ for all $c \in C$.
\begin{center}
\begin{tikzcd}
	C & A \\
	B & {(A+B)/R}
	\arrow["f", from=1-1, to=1-2]
	\arrow["g"', from=1-1, to=2-1]
	\arrow[from=1-2, to=2-2]
	\arrow[from=2-1, to=2-2]
\end{tikzcd}
\end{center}

In order to build quotient types and to define the coprojections of the pushout, we have taken a file written by Arthur Azevedo de Amorim~\cite{amorimPoleiro2024}. This implementation relies on the following standard axioms\footnote{The original file makes use of functional and propositional extensionality, but predicate extensionality happens to be sufficient.}:
\begin{itemize}
\item Predicate extensionality, which states that two pointwise equivalent predicates are equal.
\begin{lstlisting}
predicate_extensionality:
  forall A (P Q: A -> Prop), (forall\!a, P a <-> Q a) -> P = Q.
\end{lstlisting}
\item Constructive definite description, which is a form of choice, stating that, if there exists a unique object satisfying a predicate in \Prop, then we can extract a witness in \Type.
\begin{lstlisting}
constructive_definite_description :
  forall A (P: A -> Prop), (exists!x, P x) -> {x | P x}
\end{lstlisting}
\end{itemize}
It should be noted that we are building all pushouts, while the proof that a category is adhesive only requires pushouts along monos (injective functions in the case of sets). If $f$ is injective, this leads to a simplification of the relation $R$: the equivalence classes are either singletons $\{a\}$, for any $a \in A$ that is outside of the image of $f$, or contain exactly one element of $B$. This means we can have a canonical element for each equivalence class. However, computing this canonical element for elements of $A$ requires computing their inverse image under $f$, and we have not found a way to do that with a weaker system of axioms.

Predicate extensionality implies proof irrelevance, which we need to obtain pullbacks in this category (for unicity of the mediating arrow). Moreover, we use constructive definite description to prove that pushouts along monos are stable and are pullbacks. Conversely, we have shown that under proof irrelevance, the requirement of adhesive categories that all monos are regular, in \Types, is equivalent to constructive definite description.

\subsection{Setoids and Extensional Functions}
\label{ssec:setoids}

A standard technique to avoid extensionality axioms in type theory consists in replacing types with \emph{setoids}: types together with a user-defined equivalence relation. Morphisms of setoids are functions between the underlying types that map equivalent elements to equivalent elements.
We prove that the corresponding category \Setoids, where morphisms are declared equivalent when they are pointwise equivalent as functions, is adhesive.

Equipping this category with pullbacks and pushouts is easy, this time without any axiom since we get to choose how to compare elements when we define the object.
Proving that pushouts are stable and that pushouts along regular monos are pullbacks is slightly harder, but there too we do not need any axioms.
Therefore, \Setoids is rm-quasiadhesive.

To further establish that it is rm-adhesive (resp.\ adhesive) it suffices by Theorem~\ref{thm:rmadhesive} (resp.~\ref{thm:adhesive}) to prove that regular subobjects are closed under union (resp. that all monos are regular). We have proven the following equivalences:
\begin{itemize}
\item regular subobjects are closed under union in \Setoids iff \emph{disjunctions can be lifted to sums}:
  \begin{lstlisting}
forall P Q: Prop, P \/ Q -> P + Q     
  \end{lstlisting}
\item monos are regular in \Setoids iff \emph{constructive indefinite description (CID)} holds:
  \begin{lstlisting}
forall A (P: A -> Prop), (exists\!x, P x) -> {x | P x}
\end{lstlisting}
\end{itemize}
The latter property is a standard axiom; it is stronger than constructive definite description (CDD) in that it does not require unicity of the witness; the former property also follows from it (by taking $A=2$).

Therefore, while we had to use predicate extensionality and CDD from the beginning with the category \Types, we could add axioms progressively with the category \Setoids: nothing for rm-quasiadhesivity, lifting of disjunctions to sums for rm-adhesivity, and CID for adhesivity.

In particular, since the Church-Rosser and the concurrency theorems hold in rm-quasiadhesive categories, we get them in \Setoids without any axiom.

\subsection{Finite Types and Functions}
\label{ssec:fintypes}

When we restrict to finite types, we can get an adhesive category \finTypes without using any axioms.
We formalize this using the finite types and functions from the mathcomp library~\cite{mathcomp:book}, which is perfectly well-suited for our needs.

In order to factorize the code, we use the functor from \finTypes to \Setoids which maps a finite set to itself, together with propositional equality.
Being full and faithful, this functor reflects pushouts and pullbacks; it moreover preserves (regular) monos, pushouts, and pullbacks.
Therefore, we can use it to import properties like (rm-(quasi))adhesivity of \Setoids to \finTypes.

Since we want this instance to be axiom-free, we only prove rm-quasiadhesivity in this way, and we prove separately that all monos are regular in \finTypes.

We could not proceed in this way for the category \Types: CID seems to be necessary to prove that the corresponding functor preserves pushouts---even only those along regular monos. Since CID is not necessary for \Types to be adhesive, this approach would be sub-optimal. We have also proved that the categories \Types and \Setoids are equivalent, assuming predicate extensionality and CID. This combination of axioms is known to entail excluded middle in \Type, but we do not know if it is necessary for the proof of this equivalence.

\subsection{Directed Simple Graphs}
\label{ssec:sgraphs}

A \emph{directed simple graph} is a set of \emph{vertices} equipped with a binary relation of \emph{edges}. (We allow self-loops for simplicity.)
We obtain a category \SGraphs by considering \emph{graph homomorphisms}, i.e., functions preserving the edges: given a graph $(V_1, E_1)$ and a graph $(V_2, E_2)$, a morphism $f: V_1 \rightarrow V_2$ must satisfy: $\forall x,y \in V_1,\: x\: E_1\: y \Rightarrow (f\: x)\: E_2\: (f\: y)$.

Note that \Setoids is \emph{not} the subcategory of simple graphs whose edges form equivalence relations: graph homomorphisms are compared pointwise but strictly, while setoid morphisms are compared pointwise up to the target relation.

\SGraphs is rm-quasi\-adhesive~\cite{behrFundamentalsCompositionalRewriting2023};
we prove this in Rocq by reusing the fact that \Types is adhesive, and checking that the various produced functions preserve edges.

\SGraphs is not rm-adhesive, however, because the pushout operation loses information on the multiplicity of edges between two vertices.
Figure~\ref{fig:sgraph_counterex} gives a concrete counter-example: a pushout along a mono which is not a pullback.

\begin{figure}
  \centering
  \begin{tikzpicture}[scale=.7]
    \begin{scope}[every node/.style={rectangle,draw},
      every edge/.style={draw}]
        \node (A) at (0,0)
          {\begin{tikzpicture}
          \begin{scope}[every node/.style={circle,draw},
            every edge/.style={draw=black}]
              \node[style={fill=black!80}] (A1) at (0,0) {};
              \node (B1) at (1,0) {};
          \end{scope}
          \end{tikzpicture}};
        \node (B) at (4,0) 
          {\begin{tikzpicture}
          \begin{scope}[every node/.style={circle,draw},
            every edge/.style={draw=black}]
              \node[style={fill=black!80}] (A2) at (0,0) {};
              \node (B2) at (1,0) {};
           \path [-{Latex[length=2mm]}] (A2) edge (B2);
          \end{scope}
          \end{tikzpicture}};
      \node (C) at (0,-3) 
          {\begin{tikzpicture}
          \begin{scope}[every node/.style={circle,draw},
            every edge/.style={draw=black}]
              \node[style={fill=black!80}] (A3) at (0,0) {};
              \node (B3) at (1,0) {};
                    \node[style={fill=black!40}]  (C3) at (0.5,1) {};
           \path [-{Latex[length=2mm]}] (A3) edge (B3);
          \end{scope}
          \end{tikzpicture}};
      \node (D) at (4,-3) 
          {\begin{tikzpicture}
          \begin{scope}[every node/.style={circle,draw},
            every edge/.style={draw=black}]
              \node[style={fill=black!80}] (A4) at (0,0) {};
              \node (B4) at (1,0) {};
                    \node[style={fill=black!40}]  (C4) at (0.5,1) {};
           \path [-{Latex[length=2mm]}] (A4) edge (B4);
          \end{scope}
          \end{tikzpicture}};
    
      \node (E) at (-2.5,2) 
          {\begin{tikzpicture}
          \begin{scope}[every node/.style={circle,draw},
            every edge/.style={draw=black}]
              \node[style={fill=black!80}] (A4) at (0,0) {};
              \node (B4) at (1,0) {};
           \path [-{Latex[length=2mm]}] (A4) edge (B4);
          \end{scope}
          \end{tikzpicture}};
    
     \path[-{Latex[length=3mm]}] (A) edge (B);
     \path[>-{Latex[length=3mm]}] (A) edge (C);
     \path[>-{Latex[length=3mm]}] (B) edge (D);
     \path[-{Latex[length=3mm]}] (C) edge (D);
    
    \end{scope}
    \begin{scope}[every edge/.style={draw=blue}]
    \path[-{Latex[length=3mm]}] (E) edge[bend left] (B);
    \path[-{Latex[length=3mm]}] (E) edge[bend right] (C);
    \end{scope}
    \path[-{Latex[length=3mm]}] (E) edge[draw = blue, dashed] node[sloped, solid, cross out, draw=red,pos=0.5] {} (A) ;
    \node (PO) at (1.4,-1.5) {PO};
    \node[cross out, draw] (PB) at (2.4,-1.5) {PB};
  \end{tikzpicture}
  \caption{\SGraphs is not rm-adhesive.}
  \label{fig:sgraph_counterex}
\end{figure}

\subsection{Functor and Presheaf Categories}

Given categories $\C$ and $\D$, the category of functors from $\C$ to $\D$ and natural transformations has all pullbacks if $\D$ has all pullbacks, and is adhesive if $\D$ is adhesive.
The main part of the proof is to transfer pullbacks, monos and so on from natural transformations to their components and vice-versa. Proving that a natural transformation has a property if all of its components have it is generally straightforward; however the converse is often more complicated. If we have a pullback of natural transformations for example, we cannot use its universal property on one of its components. We can however use the unicity of pullbacks along with the fact that canonical pullbacks are built component-wise, but this requires having a canonical pullback in the first place. Many of those proofs thus require to have $\D$ a category with all pullbacks.

A similar issue arises for monos: we cannot directly prove that if a natural transformation is a mono then all of its components are. In this case, contrary to pullbacks, there is no notion of a category ``having all monos''. We can however use the fact that being a mono is equivalent to having a trivial kernel pair (any pullback of a mono along itself is a span of identical isomorphisms) and then use the proof for pullbacks. This is the proof from~\cite[Cor.~2.15.3]{borceuxHandbookCategoricalAlgebra1994}.
The same is probably true for regular monos, but would require a similar link with limits. Proving it would allow us to declare categories of functors as rm-(quasi)adhesive whenever $\D$ is. We leave this for future work.

Using this instance combined with the \Types (or \Setoids) instance, we deduce that the category $(\C^\op\to\Types)$ (or $(\C^\op\to\Setoids)$) of presheaves on an arbitrary category $\C$ is adhesive. This includes the category \MGraphs of directed multigraphs~\cite{behrFundamentalsCompositionalRewriting2023}, by instantiating $\C$ with the category with two objects and two parallel arrows between them. Similarly for finite presheaves $(\C^\op\to\finTypes)$, and thus finite directed multigraphs, without using any axioms.

\subsection{Slice/over Categories}

Given an object $X$ in a category $\C$, we can define the \emph{slice} category $\C\slash X$.
Its objects are the morphisms $f\colon A\to X$; a morphism from $f\colon A\to X$ to $g\colon B\to X$ is a morphism $h\colon A\to B$ in $\C$ such that $f = g \circ h$.
(Subobjects of $X$ actually form a subcategory of $\C\slash X$.)

For any instance $\C$ of a structure of the hierarchy of categories and any $X \in \C$, the slice $\C\slash X$ is an instance of the same structure as $\C$, i.e., the slice is rm-adhesive if the category is rm-adhesive, etc. As for functors and presheaf categories, the proofs mainly consist of lemmata to transfer pullbacks, pushouts, monos and regular monos to and from the slice.

This construction is a useful source of additional concrete models. For example, the category $\Types \slash X$ is the category of (intensional) multisets over $X$; and, for a chosen category of graphs, the slice category of $X$ is the category of graphs \emph{typed} by $X$, i.e. each node of $X$ defines a type and its edges define the permitted edges, so that an object of this slice category is a graph respecting the typing constraints and the morphisms between graphs are those that preserve types~\cite{monograph}. The general theory of categorical graph transformations thus lifts immediately to these richer and practically useful settings, e.g., the essential use of typing in hierarchical graph-based knowledge representation~\cite{harmer2020} and its specific application to automated schema validation and update in graph-based databases~\cite{bonifati2019}.

\section{Discussion}
\label{sec:ccl}

\subsection{Graph Transformation}

From the point of view of graph transformation, our formalization project currently provides generic, once-and-for-all proofs of the Church-Rosser and concurrency theorems for linear DPO. These proofs apply, with no further effort, to all instances of rm-quasiadhesive (or stronger) categories. In addition to some basic concrete instances---types, finite types, setoids and binary relations---we provide two generators of instances in the guise of slice categories and presheaves. As such, other key instances, such as (possibly typed, hyper-, directed) multigraphs, are already provided.

Nonetheless, there are other concrete models that we have not (yet) formalized. For example, injective functions and safely-marked Petri nets form rm-adhesive categories~\cite{johnstoneQuasitoposesQuasiadhesiveCategories2007}, and algebraic specifications form an rm-quasiadhesive category~\cite{johnstoneQuasitoposesQuasiadhesiveCategories2007}. As explained in Section~\ref{ssec:concurrency}, there are also additional theorems from graph transformation that we could formalize, a process which may lead to a strengthening of their statements, along with a precise documentation of that optimization of the literature.

\subsection{Code}

The library consists of about 10k lines of code in total. About 5100 are for the preliminaries and general category part (setoids, categories, pushouts, isomorphisms, functors, etc.), 2400 for the theory of adhesive categories (Van Kampen squares, adhesive morphisms, subobjects, rm-quasi\-adhesive categories, etc.), 400 for the Church-Rosser and concurrency theorems, and 2100 for the various instances.
The development is axiom-free, except for the proofs that the categories of types and setoids are adhesive (see Sections~\ref{ssec:types} and~\ref{ssec:setoids}---note that the category of setoids is rm-quasiadhesive without any axioms).

\subsection{Setoids and $\mathcal E$-categories}
\label{ssec:discussion:setoids}

As explained in Section~\ref{ssec:intro:equality}, we formalize $\mathcal E$-categories rather than categories: homsets are setoids. The concepts of categories scale smoothly to $\mathcal E$-categories: it suffices to
\begin{itemize}
\item replace all occurrences of propositional equality $(=)$ between morphisms by setoid equivalence $(\equiv)$
\item require that all functions between homsets preserve these equivalences (e.g., composition, or the action of a functor on morphisms).
\item define appropriate equivalence relations on morphisms whenever we define a category, e.g., compare natural transformations via their components (ignoring naturality proofs), slice morphisms via their underlying morphisms, functions in \Types pointwise, etc.
\end{itemize}
This design choice saves us from having to use functional extensionality and proof irrelevance, as well as other axioms to work with explicit quotient types.
Moreover, the latter point above regularly simplifies proofs, as equality most often \emph{reduces} to the expected property.
For instance, as explained in Remark~\ref{rmk:compcontent:setoids} we can simply define equivalence on regular or split monos to \emph{be} equivalence of the underlying morphisms, thus ignoring the additional computational content.

The price to pay is that we systematically have to prove that all concepts respect and preserve these equivalence relations.
To this end, we use the setoid rewriting tools of Rocq~\cite{Sozeau09}. For instance, we state, prove and use lemmata such as
\begin{lstlisting}
Instance isPullbackE {\C: Cat} {A B C D: \C}:
  CMorphisms.Proper (eqv==>eqv==>eqv==>eqv==>iffT)
  (@isPullback \C\_A B C D).  
\end{lstlisting}
that says that being a pullback square is closed under morphism equivalence (\lstinline{eqv}).
Proofs of such lemmata are straightforward as long as we proceed in a systematic way, and we plan to study whether tools such as Trocq~\cite{DBLP:conf/esop/CohenCM24} could help us to automate them. 

In the above example, note that since being a pullback carries computational content, it is a property in \Type, hence the use of \lstinline{CMorphisms.Proper} and \lstinline{iffT} rather than the more usual \lstinline{Proper} and \lstinline{iff} for relations in \Prop.
Rocq's support for setoid rewriting in \Type is however less robust than that in \Prop, and this currently requires some occasional unpleasant hacking if the \lstinline{rewrite} tactic fails to use the above instance automatically.

To sum up, our choice of using setoids and $\mathcal E$-categories makes it possible to avoid axioms and explicit quotients, but the price to pay is that we have to state and prove explicitly that our constructions respect the various equivalence relations we define. In contrast, a development based on propositional equality and strict categories makes it trivial to rewrite morphisms under arbitrary contexts, but would require various axioms to strengthen propositional equality (i.e., functional extensionality and proof irrelevance in general, and predicate extensionality plus choice to construct quotient types, e.g., for split and regular monos, where proof irrelevance is not an option, cf. Remark~\ref{rmk:compcontent:setoids}).

\subsection{Report on the Usability of Hierarchy Builder}
\label{ssec:hb}

HB turned out to be very convenient for the present work, and our evaluation is overall quite positive. It enables us to set up our hierarchies cleanly and efficiently, without having to tweak canonical structures or typeclass inference manually. Below we list suggestions to improve it further.

\paragraph*{Universe Polymorphism}

Universe polymorphism is not yet supported by HB, so that we had to duplicate the setoid structures in order to avoid universe inconsistencies when defining the category \Setoids: small setoids, which are the objects of this category, and large setoids, which are used for the homsets of our $\mathcal E$-categories.
Since Rocq already has support for universe polymorphic records, we do not see any major obstacles and hope that this restriction will be lifted in the near future by the developers of HB.

\paragraph*{Dynamic Hierarchies}

A notable phenomenon with our hierarchy of morphisms (Figure~\ref{fig:extended:morphisms}) is that some classes collapse when going up the hierarchy of categories. For example, in an adhesive category, every mono is adhesive (and thus regular). HB does not currently support such dynamic behavior of hierarchies. We use lemmata with adequate dependencies to promote morphisms easily when the conditions are met. Such promotions have to be explicit, however, and it would be interesting to automate the process.

\paragraph*{Inheritance a posteriori}

Our main concern is the current inability to add downward inheritance relations \emph{a posteriori}, as explained at the end of Section~\ref{ssec:vk}. Providing solutions to this problem while preserving robustness of inference seems challenging, but it is probably necessary to design and combine large libraries in a modular way.

In a setting with definitional proof irrelevance~\cite{SProp19}, it might be possible and useful to relax the forgetful inheritance policy so that it ignores mixins consisting only of propositions. Nevertheless, this would not solve our current issue with isomorphisms (Figure~\ref{fig:extended:morphisms}): the mixins for adhesive and preadhesive morphisms carry computational content.

\section{Related Work}
\label{sc:rw}

There are many formalizations of category theory in various proof
assistants~\cite{wiegleyCategoryTheoryCoq2014,leanmathlib,ahrensUnivalentCategoriesRezk2015,voevodskyUniMathComputercheckedLibrary,huFormalizingCategoryTheory2021,grossExperienceImplementingPerformant2014,starkCategoryTheoryAdjunctions2016,timanyCategoryTheoryCoq2016,vezzosi2021cubical,agda-unimath,1lab}.

Very few of these libraries define regular monos~\cite{huFormalizingCategoryTheory2021,voevodskyUniMathComputercheckedLibrary,leanmathlib}. The only formalization of adhesive categories of which we are aware is in Lean mathlib~\cite{leanmathlib}, where we find the textbook definition, the fact that pushouts along monos yield monos and are pullbacks, and the \Types and functor instances. With respect to this library, we develop more concepts (rm-quasiadhesive and rm-adhesive categories), we provide alternative presentations of theses categories (Theorems~\ref{thm:rmadhesive} and~\ref{thm:adhesive}), we prove the local Church-Rosser and Concurrency theorems, and we provide more instances (finite types, setoids, simple graphs, slices).
Another important difference is that, like in~\cite{wiegleyCategoryTheoryCoq2014,huFormalizingCategoryTheory2021}, we work with $\mathcal E$-categories to remain axiom-free, while Lean's mathlib relies on propositional extensionality and a strong form of the axiom of choice, hence classical logic---with excluded middle in \Type. In particular, these axioms allow the free taking of quotients and working with strict categories (cf. Section~\ref{ssec:intro:equality}).

\section{Future Work}
\label{sc:fw}

It would be nice to set up tools for rewriting modulo associativity of composition (we mention the associativity law 439 times in our development, in explicit rewriting sequences). While it is easy to design a tactic to prove morphism equations modulo associativity of composition and a few other basic laws of category theory, stating these equations is painful so that we would like to implement a matching algorithm to find occurrences modulo associativity and perform the appropriate transitivity steps automatically, along the lines of~\cite{BraibantP11} but for heterogeneous operations such as categorical composition.

Our library contains many lemmata related to isomorphisms: many concepts are defined only up to isomorphism, and it is often possible to introduce, remove or move isomorphisms around (e.g., lemma \lstinline{isPullback_iso_A} from Section~\ref{ssec:bundling}). Unfortunately, their use still induces increased complexity, compared to paper proofs where isomorphisms are often handled implicitly.
A typical example can be found in~\cite[pages 321-322]{johnstoneQuasitoposesQuasiadhesiveCategories2007}: ``[...] hence $B\simeq B'$. In particular, we can erase all the primes from the diagram''.
We would like to automate this task and develop a tactic for deleting a given isomorphism, keeping only its source or target object.

For instance, in a goal like
\begin{lstlisting}
forall ... A A' (i: A ≃ A') (f: A~>...) (g: ...\!\!~>A), P A f g
\end{lstlisting}
we would like to move more or less automatically to the goal
\begin{lstlisting}
forall ... A' (f: A'~>...) (g: ...\!\!~>A'), P A' f g
\end{lstlisting}
where \texttt{i} and \texttt{A} have been deleted and the types of \texttt{f} and \texttt{g} modified, using appropriate lemmata ensuring $P$ is preserved along isomorphisms.
(This is a typical case where it is far easier to work with univalent categories: isomorphisms become propositional equalities and can always be eliminated.)

Similarly, we need to develop ways of transporting results along equivalences of categories.
For example, the category of directed multigraphs can be defined either as a presheaf category (cf. Section~\ref{sec:instances}), or more concretely with any user-given definition for the type of such graphs and the type of morphisms between them. By transporting our adhesivity proof along an equivalence of categories, we would obtain for free the fact that such end-user variants are also adhesive. As mentioned in Section~\ref{ssec:duality}, there are cases where such tools would help for duality: sometimes we cannot get duality definitionally but only up to categorical equivalence.
A challenge in both cases is to be able to transport computational content in a sensible way.
The Trocq tool~\cite{DBLP:conf/esop/CohenCM24} seems promising and could even allow us to transfer results along weaker relations (e.g., adjunctions), something which would remain non-trivial even with univalence.

Finally, having a tool to display graphically the diagram currently being worked on would make the proof process easier---even just in read-only mode, unlike in the works of Ambroise Lafont~\cite{lafontDiagramEditorMechanise2024} and Luc Chabassier~\cite{chabassierGraphicalInterfaceCategory2025, chabassierAspectsCategoryTheory2025}, whose more ambitious objective is to perform proofs graphically.

\section*{Acknowledgments}
We would like to thank the anonymous reviewers for their feedback, as well as Cyril Cohen and Kazuhiko Sakaguchi for many helpful discussions on the subject of this paper and on Hierarchy Builder generally. This work was partially funded by the ANR project CoREACT: Coq-based Rewriting: towards Executable Applied Category Theory (ANR-22-CE48-0015).

\bibliographystyle{ACM-Reference-Format}
\bibliography{These}

\end{document}